\documentstyle[12pt,epsfig,epsf]{article}
\def\permil{\%\raise.10ex\hbox{$_{\scriptstyle 0}$}}

\def\beq{\begin{equation}}
\def\eeq{\end{equation}}

\newcommand{\mbf}[1]{\mbox{\boldmath $#1$}}

\newcommand{\overright}{\mbf}

\newcommand{\bk}{\mbf{k}}
\newcommand{\bq}{\mbf{q}}
\newcommand{\bp}{\mbf{p}}

\newcommand{\bx}{\mbf{x}}
\newcommand{\by}{\mbf{y}}
\newcommand{\bz}{\mbf{z}}

\newcommand{\frho}{\mbox{\boldmath $\rho$}}
\def\bea{\begin{eqnarray}}
\def\eea{\end{eqnarray}}
\begin{document}
\hfill
\hspace*{\fill}
\begin{minipage}[t]{3cm}
  DESY-04-067\\
 hep-ph/0404xxx
\end{minipage}
\vspace*{1.cm}
\begin{center}
\begin{Huge}
Interactions of Reggeized Gluons in the M\"{o}bius Representation\\[1cm]
\end{Huge}
\begin{large}
\vspace{0.5cm}
J. Bartels \\
Universit\"{a}t Hamburg\\
II. Institut f\"{u}r Theoretische Physik\\
Luruper Chaussee 149, D-22761 Hamburg, Germany;\\[1cm] 
L.N. Lipatov $^{\dagger}$ \\
Petersburg Nuclear Physics Institute\\
Gatchina, 188 300 St.Petersburg, Russia \\[1cm] 
G.P. Vacca \\
INFN Bologna, Dipartimento di Fisica \\
Via Irnerio 46, I-40126 Bologna, Italy
\end{large}
\end{center}
\vskip15.0pt \centerline{\bf Abstract}
\noindent
We investigate consequences of the M\"obius invariance of the BFKL
Hamiltonian and of the triple Pomeron vertex.
In particular, we show that the triple Pomeron vertex in QCD, when restricted
to the large $N_c$ limit and to the space of M\"obius functions, simplifies
and reduces to the vertex used in the Balitsky-Kovchegov (BK) equation.
As a result, the BK equation for the dipole density appears as
a special case of the nonlinear evolution equation which sums the
fan diagrams for BFKL Green's functions in the M\"obius representation.
We also calculate the corrections $O(1/(N_c^2-1)$ to the
triple Pomeron vertex in the space of M\"obius functions, and we present a
generalization of the BK-equation in the next-to-leading order approximation 
in the $1/N_c$ expansion.
\vskip 1cm \hrule
\vskip 1cm \noindent \noindent $^{(\dagger)}$ {\it Humboldt Preistr\"ager
\newline
Work supported in part by INTAS and by the Russian Fund of Fundamental
Investigations} 
\vfill
%%%%%%%%%%%%%%%%%%%%%%%%%%%%%%%%%%%%%%%%%%%%%%%%%%%%%%%%%%%%%%%%%%%%%%%%%%%%
\section{Introduction}

It has been observed long time ago that the LO kernel of the BFKL equation
~\cite{BFKL} is invariant under M\"obius transformations ~\cite{conf}.
The same
invariance holds for the transition vertex of $2 \to 4$ reggeized gluons
~\cite{Bartels:1995kf}. This symmetry, together with the fact that,
in physical scattering
processes, the Green's functions of reggeized gluons couple to impact factors
of colorless projectiles which vanish as the momentum of any of the
attached reggeized gluons goes to zero,
leads to a freedom of redefining the Green's functions of
reggeized gluons. In particular, the Pomeron Green's functions of two
gluons in configuration space, $f(\frho_1, \frho_2)$,
can be redefined to have the property $f(\frho,\frho)=0$. Functions of this
type will be named as `being in the M\"obius representation' or,
alternatively, as `belonging to the M\"obius space of functions'.

In this paper we investigate some consequences of this M\"obius
representation for the interactions of reggeized gluons. After a brief
review of the M\"obius representation of solutions to the BFKL equation,
we investigate the connection between the reggeon Green's functions
in the M\"obius representation and the dipole picture ~\cite{dipole}.
We then turn to the
BKP equations ~\cite{BKP}, and, finally, to the nonlinear equation of
fan diagrams
which, in the large $N_c$ limit, is shown to coincide with the
Balitsky-Kovchegov equation ~\cite{BalKov}. This equation is currently
intensively studied in connection with saturation, and it has 
been rederived in the framework of different approaches \cite{BJIMWLK}.

We also compute the contribution of the non planar part of the
transition vertex of $2 \to 4$ reggeized gluons,
which is subleading in $1/N_c$ and leads to a
new contribution in the nonlinear equation for fan diagrams.
We construct a system of coupled equations in the
next-to-leading order approximation of the $1/N_c$ expansion.
%%%%%%%%%%%%%%%%%%%%%%%%%%%%%%%%%%%%%%%%%%%%%%%%%%%%%%%%%%%%%%%%%%%%%%%%%%
\section{Review of the BFKL Hamiltonian}

To provide the conformal invariance
of the BFKL equation \cite{BFKL} initially one should
substitute the Born $t$-channel partial wave $%
f_{\omega }^{Born}$ ($\omega =j-1$), which is proportional to the product of
two gluon Green functions $\ln \,\left| \rho _{11^{\prime }}\right| ^{2}$ and $\ln
\,\left| \rho _{22^{\prime }}\right| ^{2}$,
by the function $f_{\omega}^{(0)} $ of anharmonic ratios \cite{conf}
\begin{equation}
\frac{1}{\omega }\ln \,\left| \rho _{11^{\prime }}\right| ^{2}\,\ln \,\left|
\rho _{22^{\prime }}\right| ^{2}\rightarrow f_{\omega }^{(0)}(%
\overright{\rho }_{1},\overright{\rho }_{2},\overright{\rho }%
_{1^{\prime }},\overright{\rho }_{2^{\prime }})\,,\,\,
\end{equation}
\begin{equation}
f_{\omega }^{(0)}(\overright{\rho }_{1},\overright{\rho }_{2};%
\overright{\rho }_{1^{\prime }},\overright{\rho }_{2^{\prime }})=%
\frac{2}{\omega }\,\ln \,\frac{\left| \rho _{11^{\prime }}\right| \left|
\rho _{22^{\prime }}\right| }{\left| \rho _{12^{\prime }}\right| \left| \rho
_{21^{\prime }}\right| }\,\ln \,\frac{\left| \rho _{11^{\prime }}\right|
\left| \rho _{22^{\prime }}\right| }{\left| \rho _{12}\right| \left| \rho
_{1^{\prime }2^{\prime }}\right| },
\end{equation}
which is invariant under the M\"obius transformation
\begin{equation}
\rho _{k}\rightarrow \frac{a\rho _{k}+b}{c\rho _{k}+d}
\end{equation}
for arbitrary complex parameters $a,b,c$ and $d$. We used here the complex
coordinates for the initial ($k=1,2$) and final ($k=1^{\prime },2^{\prime }$%
) gluons in the two-dimensional impact parameter
space $\overright{\rho }=(x,y)$

\begin{equation}
\rho _{k}=x_{k}+iy_{k}\,,\,\,\rho _{k}^{\ast }=x_{k}-iy_{k}\,,\,\,\,\rho
_{kl}=\rho _{k}-\rho _{l}\,.
\end{equation}

Such a substitution can be justified by making use of the fact that the impact
factors $\Phi ^{(A,B)}$ of colliding colourless particles
\beq
\Phi _{q}^{A}(\overright{\rho }_{1},\overright{\rho }_{2})=e^{i
\overright{q} (\overright{\rho }_{1}+\overright{\rho}
_{2})/2}\int \frac{d^{2}k\,}{(2\pi )^{2}}\,e^{i\overright{k}
\overright{\rho }_{12}}\,\Phi ^{A}(\frac{\overright{q}}{2}+%
\overright{k},\frac{\overright{q}}{2}-\overright{k})\,,\,\,%
\overright{k}=\frac{\overright{k}_{1}-\overright{k}_{2}}{2}%
\,,
\eeq
\beq
\Phi _{q^{\prime }}^{B}(\overright{\rho }_{1^{\prime }},\overright{%
\rho }_{2^{\prime }})=e^{-i\overright{q^{\prime }} (\overright{
\rho }_{1^{\prime }}+\overright{\rho }_{2^{\prime }})/2}\int \frac{%
d^{2}k\,}{(2\pi )^{2}}\,e^{-i\overright{k} \overright{\rho }
_{1^{\prime }2^{\prime }}}\,\Phi ^{A}(\frac{\overright{q^{\prime }}}{2}%
+\overright{k},\frac{\overright{q^{\prime }}}{2}-\overright{k}%
)\,,\,\,\overright{k}=\frac{\overright{k}_{1}-\overright{k}%
_{2}}{2}\,.
\eeq
entering in the expression for the scattering amplitude at high energies
$\sqrt{s}$ and fixed $t=-\overright{q}^{2}$ in the leading logarithmic
approximation (LLA).
\begin{equation}
A(s,t)=is\int_{\sigma -i\infty }^{\sigma +i\infty }\frac{d\omega }{2\pi i}%
\,\,e^{\omega Y}\,\,f_{\omega }(t)\,,\,Y=\ln \,s\,,
\end{equation}
\begin{equation}
\,T_{\omega }^{A,B}=f_{\omega }(t)\,\delta ^{2}(q-q^{\prime })=\int
\prod_{r=1,2,1^{\prime },2^{\prime }}d^{2}\rho _{r}\,\,\,\Phi _{q}^{A}(%
\overright{\rho }_{1},\overright{\rho }_{2})\,f_{\omega }(%
\overright{\rho }_{1},\overright{\rho }_{2};\overright{\rho }%
_{1^{\prime }},\overright{\rho }_{2^{\prime }})\,\,\Phi _{q^{\prime
}}^{B}(\overright{\rho }_{1^{\prime }},\overright{\rho }%
_{2^{\prime }})
\label{amplitude}
\end{equation}
satisfy the relations \cite{conf}
\begin{equation}
\int d^{2}\rho _{1}\,\Phi _{q,q^{\prime }}^{A,B}(\overright{\rho }_{1},%
\overright{\rho }_{2})=\int d^{2}\rho _{2}\,\Phi _{q,q^{\prime }}^{A,B}(%
\overright{\rho }_{1},\overright{\rho }_{2})=0\,.
\end{equation}
These relations are consequences of gauge invariance of the impact
factors. In the momentum representation they simply read as follows:
\begin{equation}
\Phi ^{(A,B)}(0,\overright{k}_{2})=\Phi ^{(A,B)}(\overright{k}%
_{1},0)=0
\end{equation}
The last property has the interpretation that
the interaction of a gluon with a small
transverse
momentum is proportional to the vanishing total colour charge of the
colliding particle.

Note, that in an accordance with Ref. \cite{conf} the partial wave $%
T_{\omega }^{A,B}$ includes the $\delta $-function $\delta ^{2}(q-q^{\prime
})$ corresponding to the momentum conservation in the cross channel, and
the impact factors $\Phi ^{(A,B)}(\frho_1,\frho_2)$ in the coordinate
representation are obtained from the
impact factors $\Phi ^{(A,B)}(\bk_1,\bk_2)$\ in the momentum representation
by the Fourier transformation.

The partial wave $f_{\omega }(\overright{\rho }_{1},\overright{%
\rho }_{2};\overright{\rho }_{1^{\prime }},\overright{\rho }%
_{2^{\prime }})$ $\ $for the gluon-gluon scattering in LLA satisfies the
BFKL equation \cite{BFKL}:

\begin{equation}
f_{\omega }(\overright{\rho }_{1},\overright{\rho }_{2};%
\overright{\rho }_{1^{\prime }},\overright{\rho }_{2^{\prime
}})=f_{\omega }^{(0)}(\overright{\rho }_{1},\overright{\rho }_{2};%
\overright{\rho }_{1^{\prime }},\overright{\rho }_{2^{\prime }})-
\frac{\bar{\alpha}_s}{2 \omega }\,H_{12}\,f_{\omega }(\overright{\rho }%
_{1},\overright{\rho }_{2};\overright{\rho }_{1^{\prime }},%
\overright{\rho }_{2^{\prime }})\,,
\end{equation}
where $\bar{\alpha}_s=\alpha_s N_{c}/\pi$.
The BFKL Hamiltonian $H_{12}$ in the leading logarithmic approximation (LLA)
can be written in the operator form (see \cite{integrab}):

\begin{equation}
H_{12}=\ln \,\left| p_{1}\right| ^{2}+\ln \,\left| p_{2}\right| ^{2}+\frac{1%
}{p_{1}p_{2}^{\ast }}\ln \,\left| \rho _{12}\right| ^{2}\,p_{1}p_{2}^{\ast
}\,+\frac{1}{p_{1}^{\ast }p_{2}}\ln \,\left| \rho _{12}\right|
^{2}\,p_{1}^{\ast }p_{2}-4\Psi (1)\,,
\label{BFKL_mom}
\end{equation}
where $\Psi (x)=d\ln \Gamma (x)/dx$, and we introduced the gluon
holomorphic momenta

\begin{equation}
p_{r}=i\frac{\partial }{\partial \rho _{r}}\,,\,\,p_{r}^{\ast }=i\frac{%
\partial }{\partial \rho _{r}^{\ast }}\,.
\end{equation}

It is important that the BFKL equation is implied to be projected to the
class of functions $\widetilde{\Phi }(\overright{p}_{1},\overright{%
p}_{2})$ vanishing at $\overright{p}_{1}=0$ and $\overright{p}%
_{2}=0$ \cite{conf}. Because the function
\beq
\widetilde{\widetilde{\Phi }}(\overright{p}_{1},\overright{p}_{2})=%
\widetilde{\Phi }(\overright{p}_{1},\overright{p}_{2})H_{12}
\eeq
also has the same properties, we conclude that the solution of the
homogeneous BFKL equation

\beq
E\,f(\overright{\rho }_{1},\overright{\rho }_{2})=H_{12}\,\,f(%
\overright{\rho }_{1},\overright{\rho }_{2})\,,\,\,E=-\,\frac{2}
{\bar{\alpha}_s}\,\omega \,.\,
\eeq
is invariant under the substitution

\begin{equation}
f(\overright{\rho }_{1},\overright{\rho }_{2})\rightarrow
\tilde{f}(\overright{\rho }_{1},\overright{\rho }_{2}) =
f(\overright{\rho }_{1},\overright{\rho }_{2})+
f^{(1)}(\overright{\rho }_{1})+f^{(2)}(\overright{\rho }_{2})\,,
\label{uv_transf}
\end{equation}
where $f^{(r)}(\overright{\rho }_{r})$ are arbitrary functions. One can
use this freedom to impose the additional constraint on $f_{\omega }(%
\overright{\rho }_{1},\overright{\rho }_{2})$

\begin{equation}
\tilde{f}(\overright{\rho },\overright{\rho })=0\, ,
\end{equation}
by choosing $f^{(i)}(\overright{\rho }_{i})=-1/2
f(\overright{\rho }_{i},\overright{\rho }_{i})$.
We define the solutions having this property as functions belonging to the
M\"{o}bius representation. This definition is in accordance with the fact
that in such a class of functions the homogeneous BFKL equation is
invariant under M\"{o}bius transformations. \
Moreover, the conformal symmetry gives a possibility to find its solutions
\cite{conf} in the form:

\begin{equation}
E_{m,\widetilde{m}}(\overright{\rho }_{1},\overright{\rho }%
_{2})=\left( \frac{\rho _{12}}{\rho _{10}\rho _{20}}\right) ^{m}\left( \frac{%
\rho _{12}^{\ast }}{\rho _{10}^{\ast }\rho _{20}^{\ast }}\right) ^{%
\widetilde{m}}
\end{equation}
where the conformal weights $m$ and $\widetilde{m}$ are equal to

\beq
m=\frac{1}{2}+i\nu +\frac{n}{2}\,,\,\,\widetilde{m}=\frac{1}{2}+i\nu -\frac{n%
}{2}
\eeq
for the principal series of unitary representations. They parametrize the
eigenvalues of two Casimir operators of the global conformal group

\beq
M^{2}\,f_{m,\widetilde{m}}=m(m-1)\,f_{m,\widetilde{m}}\,,\,\,M^{\ast
2}\,f_{m,\widetilde{m}}=\widetilde{m}(\widetilde{m}-1)\,\,f_{m,\widetilde{m}%
}\,,
\eeq
where

\beq
M^{2}=\left( \sum_{r=1}^{2}\overright{M}_{r}\right) ^{2}=2\left(
\overright{M}_{1},\overright{M}_{2}\right) =-\rho
_{12}^{2}\partial _{1}\partial _{2}\,,\,\,\partial _{r}=\frac{\partial }{%
\partial \rho _{r}}\,.
\eeq
Here $\overright{M}_{r}$ are the generators of the M\"{o}bius group
\beq
M_{r}^{3}=\rho _{r}\partial _{r}\,,\,\,M_{r}^{+}=\partial
_{r}\,,\,\,M_{r}^{-}=-\rho _{r}\partial _{r}\,.
\eeq
If we chose $f_{\omega }^{(0)}(\overright{\rho }_{1},\overright{%
\rho }_{2};\overright{\rho }_{1^{\prime }},\overright{\rho }%
_{2^{\prime }})$ as an inhomogeneous term of the BFKL equation, its solution
is also conformally invariant \cite{conf} because the iteration of $\
f_{\omega }^{(0)}$ always gives functions belonging to the M\"{o}bius
representation.

The BFKL Hamiltonian has the property of the holomorphic separability,~\cite
{separab}

\beq
H_{12}=h_{12}+h_{12}^{\ast },\,\,h_{12}=\sum_{r=1}^{2}\left( \ln p_{r}+\frac{%
1}{p_{r}}\ln (\rho _{12})\,p_{r}-\Psi (1)\right) \,.
\label{BFKL_sep}
\eeq
This representation is valid in the space of M\"obius functions, where terms
proportional to $\delta^{(2)}(\rho_{12})$ (arising from
$\nabla^2 \log{|\rho|} = 2\pi \delta^{(2)}(\rho)$) can be neglected.
Alternatively, the holomorphic Hamiltonian $h_{12}$ can be written in
another operator form \cite{integrab}
\beq
h_{12}=\sum_{r=1}^{2}\left( \ln \rho _{12}+\rho _{12}\ln (p_{r})\frac{1}{%
\rho _{12}}-\Psi (1)\right) \,,
\label{BFKL_sep_dual}
\eeq
where we have used the relations
\beq
\log p = \frac{i}{\rho \,p} + \rho (\log{p}) \, \frac{1}{\rho} \quad ,
\quad
\frac{1}{p} (\log{\rho}) \, p = - \frac{i}{\rho \,p} + \log \rho \,.
\label{dual_rel}
\eeq
It means, that the total Hamiltonian $H_{12}$ can be presented as follows

\beq
H_{12}=2\ln \,\left| \rho _{12}\right| ^{2}+\left| \rho _{12}\right| ^{2}\ln
\,\left| p_{1}p_{2}\right| ^{2}\,\,\frac{1}{\left| \rho _{12}\right| ^{2}}%
\,-4\,\Psi (1)\,,
\label{BFKL_dual}
\eeq
where terms proportional to $\delta^{(2)}(\mbf{p}_i)$ have been
neglected, since physical amplitudes have to be integrated with colourless
impact factors. Finally, in the M\"{o}bius representation the Hamiltonian
of the BFKL equation
can be written as the integral operator:
\begin{equation}
H_{12}\,f_{\omega }(\overright{\rho }_{1},\overright{\rho }%
_{2})=\int \frac{d^{2}\rho _{3}}{\pi }\,\frac{\left| \rho _{12}\right| ^{2}}{%
\left| \rho _{13}\right| ^{2}\left| \rho _{23}\right| ^{2}}\,\left(
f_{\omega }(\overright{\rho }_{1},\overright{\rho }_{2})-f_{\omega
}(\overright{\rho }_{1},\overright{\rho }_{3})-f_{\omega }(%
\overright{\rho }_{2},\overright{\rho }_{3})\right)\,.
\end{equation}
Indeed, by introducing an intermediate ultraviolet regularization with $%
\delta \rightarrow 0$, we reproduce $H_{12}$ in the above operator form,
because

\beq
\int \frac{d^{2}\rho _{3}}{\pi }\,\frac{\left| \rho _{12}\right| ^{2}}{%
\left( \left| \rho _{13}\right| ^{2}+\delta ^{2}\right) \left( \left| \rho
_{23}\right| ^{2}+\delta ^{2}\right) }=\int_{0}^{1}dx\frac{\left| \rho
_{12}\right| ^{2}}{x(1-x)\left| \rho _{12}\right| ^{2}+\delta ^{2}}\simeq
2\ln \,\frac{\left| \rho _{12}\right| ^{2}}{\delta ^{2}}\,,
\label{dipole_virt}
\eeq
\begin{eqnarray}
\!&-&\!\left| \rho _{12}\right| ^{2}\int
\frac{d^{2}\rho _{3}}{\pi \left( \left|
\rho _{23}\right| ^{2}+\delta ^{2}\right) }\,\frac{\,f_{\omega }(%
\overright{\rho }_{1},\overright{\rho }_{3})}{\left| \rho
_{13}\right| ^{2}}\,\simeq \left| \rho _{12}\right| ^{2}\left( \ln \,(\delta
^{2}\left| p_{2}\right| ^{2})-\,2\Psi (1)\right) \,\frac{\,f_{\omega }(%
\overright{\rho }_{1},\overright{\rho }_{2})}{\left| \rho
_{12}\right| ^{2}}\,\,, \nonumber \\
\!&-&\!\left| \rho _{12}\right| ^{2}\int
\frac{d^{2}\rho _{3}}{\pi \left( \left|
\rho _{13}\right| ^{2}+\delta ^{2}\right) }\,\frac{f_{\omega }(%
\overright{\rho }_{2},\overright{\rho }_{3})}{\left| \rho
_{23}\right| ^{2}}\,\simeq \left| \rho _{12}\right| ^{2}\left( \ln \,(\delta
^{2}\left| p_{1}\right| ^{2})-\,2\Psi (1)\right) \,\frac{\,f_{\omega }(%
\overright{\rho }_{1},\overright{\rho }_{2})}{\left| \rho
_{12}\right| ^{2}}\,.
\label{dipole_real}
\end{eqnarray}

In this form, the BFKL Hamiltonian was presented first in the context of the
dipole picture \cite{dipole} (see also \cite{BalKov}).
It is instructive to trace, following the path of transformations from eq.
(\ref{BFKL_mom}) and (\ref{BFKL_sep}) to (\ref{BFKL_sep_dual}) and
(\ref{BFKL_dual}), the gluon reggeization and the real production terms
(which are connected to each other due to the
bootstrap relation).
Starting from (\ref{BFKL_mom}), (\ref{BFKL_sep}),
we consider the terms related to the reggeized gluon trajectory,
$\log{|p_r|^2}$. The use of the relation eq. (\ref{dual_rel})
takes us to the form $\left| \rho _{12}\right| ^{2}\ln \,\left| p_{r}
\right| ^{2}\,\,\frac{1}{\left| \rho _{12}\right| ^{2}}$ (apart from the terms
$\frac{1}{\rho_{12}p_r}$ which cancel when combined with the corresponding 
terms from the real production). 
When applying the relation (\ref{dipole_real}),
these terms are identified with those pieces which, in the dipole approach,
are obtained from the real production.
In the same way, those terms which in (\ref{BFKL_mom}), (\ref{BFKL_sep})
are associated with the real production $\frac{1%
}{p_{1}p_{2}^{\ast }}\ln \,\left| \rho _{12}\right| ^{2}\,p_{1}p_{2}^{\ast
}\,+\frac{1}{p_{1}^{\ast }p_{2}}\ln \,\left| \rho _{12}\right|
^{2}\,p_{1}^{\ast }p_{2}$ are transformed into the $2\ln \,\left| \rho _{12}\right|
^{2}$ term (taking into account the cancellation mentioned previously).
Finally, thanks to the relation in eq. (\ref{dipole_virt}), one
finds that this term gives the virtual (one loop) contribution in the
dipole picture.
Thus, when going from the momentum space representation of the
BFKL Hamiltonian to the dipole picture, one observes a (partial)
exchange of the virtual and real contributions and of the U.V. and I.R.
sectors, which corresponds to the duality transformation \cite{dual}.
%%%%%%%%%%%%%%%%%%%%%%%%%%%%%%%%%%%%%%%%%%%%%%%%%%%%%%%%%%%%%%%%%%%%%%%%%%%%
\section{The M\"{o}bius representation and the dipole picture}

In the dipole approach \cite{dipole} one introduces the dipole distribution
in a hadron as a function of the rapidity $Y=\ln \,s$,
$N_{\overright{\rho }_{1},\overright{\rho }_{2}}$.
The BFKL equation is written in the form:
\beq
\frac{dN_{\overright{\rho }_{1},\overright{\rho }_{2}}}{dY}=-\,%
\frac{\bar{\alpha}_s}{2 }\int \frac{d^{2}\rho _{3}}{\pi }\,\frac{\left|
\rho _{12}\right| ^{2}}{\left| \rho _{13}\right| ^{2}\left| \rho
_{23}\right| ^{2}}\,\left( N_{\overright{\rho }_{1},\overright{%
\rho }_{2}}-N_{\overright{\rho }_{1},\overright{\rho }_{3}}-N_{%
\overright{\rho }_{2},\overright{\rho }_{3}}\right) .
\eeq
$N_{\frho_1,\frho_2}$ can be interpreted as the scattering amplitude
of a color dipole ( e.g. a quark-antiquark pair).
It gives the possibility to calculate the total cross-sections
$\sigma _{t}$ at high energies $\sqrt{s}$:
\beq
\sigma _{t}=\int d^{2}\rho _{1}d^{2}\rho _{2}\,\int_{0}^{1}dx\,\left| \psi
_{p}(\frho_1, \frho_2)\right| ^{2}\,N_{\overright{\rho }%
_{1},\overright{\rho }_{2}}(Y)\,.
\eeq
Here $\psi _{p}(\overright{\rho }_{1},\overright{%
\rho }_{2};x)$ denotes the wave function of the colourless state of
the projectile, being a composite state of two quarks with transverse
coordinates $\overright{\rho }_{1}$,
$\overright{\rho }_{2}$ and longitudinal momentum fractions $x$, $1-x$.

In order to illustrate the connection ~\cite{NavWall} between this cross section formula and
the discussion presented above, we consider,
as an example, the elastic scattering of two virtual photons with momentum
transfer squared $t=-\bq^2$.
In leading order, the impact factor $\Phi^{\gamma^*}$ is simply given by a
closed quark loop with the $t$-channel gluons being attached in all possible
ways. Starting from Feynman diagrams in momentum space und taking suitable
Fourier transforms one obtains the following form for the
scattering amplitude (\ref{amplitude}) ~\cite{BGP}:
\begin{eqnarray}
T^{\gamma^*\gamma^*} = is \int d^2 \rho_1 d^2 \rho_2 d^2 \rho_1' d^2 \rho_2'
e^{i\bq (\frho_1 + \frho_2)/2} \int_0^1 dx
\psi_{p_2}(\frho_{12},x)^* \psi_{p_1} (\frho_{12}, x)
\nonumber\\
\tilde{G} (\frho_{1},\frho_{2}; \frho_{1'},\frho_{2'};Y)
\nonumber\\
e^{-i\bq (\frho_1' + \frho_2')/2} \int_0^1 dx' \psi_{p_2'}(\frho_{1'2'},x')^*
\psi_{p_1'} (\frho_{1'2'}, x')\,,
\end{eqnarray}
where $p_1$, $p_{1'}$ ($p_2$, $p_{2'}$) denote the transverse momenta
of the incoming (outgoing) photons, $x$ and $x'$ are the longitudinal
momentum fractions inside the impact factors.
$\psi_p$ is the wave function  of the virtual photon with transverse
momentum $\bp$. Its dependence on the transverse momentum is contained in
a phase factor:
\beq
\psi_p(\frho, x) = \psi(\frho, x) e^{i(1-x) \bp \frho_{12}},
\eeq
where $\psi(\frho, x)$ denotes the usual photon wave function
used in the total cross section formula. $\tilde{G}$ stands for the
following Fourier transform of the momentum space BFKL Green's function of
reggeized gluons, $G_{\omega}(\bk_1,\bk_2;\bk_1',\bk_2';Y)$:
\begin{eqnarray}
\tilde{G} (\frho_{1},\frho_{2}\frho_{1'}\frho_{2'};Y)=
\int d^2 k d^2 k' e^{i\bk \frho_{12}}
\left( 1 - e^{-i (\bk + \frac{\bq}{2}) \frho_{12}} \right)
     \left( 1 - e^{i (-\bk + \frac{\bq}{2}) \frho_{12}} \right)\nonumber \\
      G (\bk+\frac{\bq}{2},-\bk+\frac{\bq}{2};
                 \bk'+\frac{\bq}{2},-\bk'+\frac{\bq}{2};Y)\nonumber \\
      e^{-i\bk \frho_{1'2'}}
     \left( 1 - e^{i (\bk + \frac{\bq}{2}) \frho_{1'2'}} \right)
      \left( 1 - e^{-i (-\bk + \frac{\bq}{2}) \frho_{1'2'}} \right)\,.
\label{modifiedG}
\end{eqnarray}
This leads to the following identification:
\begin{eqnarray}
N_{\frho_1,\frho_2} = \int d^2 \rho_1' d^2 \rho_2'
\tilde{G} (\frho_{1},\frho_{2}; \frho_{1'},\frho_{2'};Y)
e^{-i\bq (\frho_1' + \frho_2')/2} \int_0^1 dx' \psi_{p_2'}(\frho_{1'2'},x')^*
\psi_{p_1'} (\frho_{1'2'}, x')\,.
\end{eqnarray}
In particular, the dipole scattering amplitude  $N_{\frho_1,\frho_2}$
is not simply the Fourier transform of the momentum space Green's function
but contains extra phase factors written in (\ref{modifiedG}). These factors garantee
that $N_{\frho_1,\frho_2}$ vanishes as $\frho_{12}$ tends to zero.

Another way to see how the gauge freedom allows us move from one
representation to the other can be summarized in the following way,
which will be useful in the study of the resummed fan diagram structure.
Let us call $\theta_{IR}$ the collection of phase factors which, in the impact
factor of a photon which splits in a $q\bar q$ pair, ties the squared
modulus of the wave function to the Green's function 
(in eq.(\ref{modifiedG}), $\theta_{IR}$ stands for the phase factors in the
first line (upper impact factor) or in the lower line (lower impact factor)): 
these $\theta_{IR}$ factors are zero if one of the two gluon momenta 
vanishes, and it contains subtraction terms with a
$\delta^{(2)}(\rho_{12})$ behavior in the coordinate representation.
We also introduce the operator $\theta^{UV}$, related to the transformation
introduced in eq. (\ref{uv_transf}) which contains terms
proportional to $\delta^{(2)}(\mbf{p}_i)$.
Using a shorthand notation and omitting the spatial integrations, one may 
write:
\beq
\Phi G= |\psi|^2 \theta_{IR} G=
|\psi|^2 \theta_{IR} \theta^{UV}G=
|\psi|^2 \theta^{UV}G=|\psi|^2 \tilde{G} \, ,
\label{subtraction}
\eeq
where $\theta^{UV}$ is chosen in order to kill the subtractions
contained in $\theta_{IR}$.

Results of this example can easily be generalized.   
It is possible to prove that the solution of the Bethe-Salpeter equation
for the Pomeron wave function $f(\overright{\rho }_{1},\overright{%
\rho }_{2};Y)$ in the M\"{o}bius representation coincides with the dipole
distribution $N_{\overright{\rho }_{1},\overright{\rho }_{2}}(Y)$.
Both functions satisfy the same BFKL equation, and they
vanish at $\overright{\rho }_{1}=\overright{\rho }_{2}$.
An advantage of using the Pomeron wave function
$f(\overright{\rho }_{1},\overright{\rho }_{2};Y)$
in the M\"{o}bius representation lies in the fact
that the amplitude for the scattering of colorless particles is
expressed as a convolution of the impact factors $\Phi _{q}^{A,B}(%
\overright{\rho }_{1},\overright{\rho }_{2})$ and the Green
function $f$ for reggeized gluon interactions. The vanishing
of\thinspace\ $f(\overright{\rho }_{1},\overright{\rho }_{2};Y)$
at $\overright{\rho }_{1}=\overright{\rho }_{2}$ means that, when
performing the integration over
$\overright{\rho }_{1}$ and $\overright{\rho }_{2}$,
in the impact factor $\Phi _{q}(\overright{\rho }_{1},%
\overright{\rho }_{2})$ we can omit the terms proportional to
$\delta ^{2}(\rho_{12})$:
these contributions correspond to those Feynman diagrams
where the reggeized gluons are attached to the same quark or gluon line.
As to the remaining Feynman diagrams in which the gluons are
attached to different lines, their contributions can be expressed in
terms of a colour density matrix
$\Omega _{p_{1}p_{1^{\prime }}}(\overright{\rho }_{1},%
\overright{\rho }_{2})$

\beq
\Phi _{q}(\overright{\rho },\overright{\rho ^{\prime }}%
)\longrightarrow \Omega _{p_{1}p_{1^{\prime }}}(\overright{\rho },%
\overright{\rho ^{\prime }})=
\eeq
\beq
e^{i(\overright{\rho }+\overright{\rho ^{\prime }})(\overright{%
p}_{1^{\prime }}-\overright{p}_{1})/2}\sum_{n}\int \prod_{k=1}^{n}dx_{k}%
\frac{d^{2}\rho _{k}}{2\pi }\,\,\psi _{p_{1^{\prime }}}^{\ast }\delta
(1-\sum_{k=1}^{n}x_{k})\sum_{i\neq l}T_{i}^{a}T_{l}^{a}\,\delta ^{2}(\rho
_{i}-\rho )\,\delta ^{2}(\rho _{l}-\rho ^{\prime })\,\psi _{p_{1}}\,\,.
\eeq
The wave functions $\psi _{p_{1}}$ and $\psi _{p_{1^{\prime }}}$ of
the initial and final colourless particles contain the Fock states with an
arbitrary number $n$ of gluons and quarks with longitudinal momenta $%
px_{k}$ and transverse coordinates $\overright{\rho }_{k}$. Due to the
translational invariance they depend only on differences of $\overright{%
\rho }_{k}$.

The wave functions $\psi _{p_{r}}$ contain also a dependence on colour
degrees of freedom of gluons and quarks. It is implied that the colour
group generator $T_{i}^{a}$ acts on the colour indices of the parton $i$ and
belongs to the corresponding representation of the colour group algebra $%
SU(N_{c})$. It means that only in the large-$N_{c}$ limit, where
in color space the gluons can be visualized as being composite
quark-antiquark states, the color density matrix is reduced to the dipole
density. Note that, in
general, the integrals over the variables $x_{k}$ are divergent at small
values and should be regularized in order to avoid double-counting. Indeed,
for the case of gluons the integration over the small-$x$ region is taken
already into account in the BFKL resummation.

In LLA it is natural to leave in the parton wave
functions for initial and final particles only the quark-antiquark component

\beq
\psi _{p_{r}}(\overright{\rho }_{1},\overright{\rho }_{2};x)=e^{i(%
\overright{\rho }_{1}+\overright{\rho }_{2},\overright{p}%
_{r})/2}\psi _{p_{r}}(\overright{\rho }_{12};x)\,.
\eeq
Then the color density matrix is simplified
\beq
\Omega _{p_{1}p_{1^{\prime }}}(\overright{\rho },\overright{\rho
^{\prime }})=e^{i(\overright{\rho }+\overright{\rho ^{\prime }})
(\overright{p}_{1^{\prime }}-\overright{p}_{1})/2}\,\frac{
N_{c}^{2}-1}{2N_{c}}\int_{0}^{1}dx\,\,\psi _{p_{2}}^{\ast }(\overright{%
\rho }-\overright{\rho ^{\prime }};x)\,\psi _{p_{1}}(\overright{%
\rho }-\overright{\rho ^{\prime }};x)\,.
\eeq
Therefore we obtain the dipole expressions discussed before.

Thus, in the M\"{o}bius representation both the reggeon and the dipole
pictures
are compatible with each other. In particular, in the reggeon language the
fact that those diagrams where both reggeized gluons
are attached to the same quark or gluon line (impulse approximation)
give a vanishing contribution makes it natural for the BFKL Pomeron
to be viewed as a Mandelstam cut. Indeed, for the Mandelstam cut the impact
factor should contain only contributions of non-planar diagrams with
non-zero third spectral function $\rho (s_{1},u_{1})$, whereas the diagrams of
the impulse approximation do not contain singularities in one of two
channels $s_{1}$ or $u_{1}$. Moreover, in QCD we can use the space-time
picture for visualizing the Mandelstam cut as describing two independent
parton fluctuations, produced by the high energy initial particle
at $t\rightarrow -\infty$, long before the collision.
The two fluctuations consist of large numbers of gluons
which in the rapidity interval $0\leq y\leq \ln \,s$ are
distributed homogeneously. The softest partons of each fluctuation
interact simultaneously with the target.
In this picture one can calculate not only the behaviour of
total cross-sections, but, taking into account the AGK cutting rules,
also the distribution of the produced particles inside the BFKL Pomerons.
An important advantage of the reggeon approach over the dipole picture
is the possibility of taking into account the non-trivial phase structure
of reggeon diagrams related to their signature factors. The scattering
amplitudes constructed in the framework of the reggeon calculus in QCD will
satisfy the requirements of the $t$- and $s$- channel unitarities. In
particular, $s$-channel unitarity is incorporated partly in the bootstrap
relations for reggeon diagrams. In the dipole approach, both the bootstrap
properties and $t$-channel unitarity remain somewhat obscure.

%%%%%%%%%%%%%%%%%%%%%%%%%%%%%%%%%%%%%%%%%%%%%%%%%%%%%%%%%%%%%%%%%%%%%%%
\section{The BKP equations in the M\"{o}bius representation}

The BFKL approach can be generalized to the case of colourless composite
states constructed from $n$ reggeized gluons \cite{BKP}. The homogeneous BKP
equation for the $t$-channel partial wave in LLA has the form (see
\cite{integrab})

\beq
Ef=Hf\,,\,\,H=\sum_{1\leq k\leq l\leq n}\frac{T_{k}^{a}T_{l}^{a}}{(-N_{c})}%
H_{kl}\,,\,E=-\frac{2\pi }{\alpha _{c}N_{c}}\omega \,,
\eeq
where $T_{k}^{a}$ are the colour group generators acting on colour indices
of the gluon $k$. The spectrum of energies $E$ allows one to find the
intercepts $\omega $ of the colourless reggeon bound states governing the
corresponding contribution $\sim s^{\omega }$ to the total cross-section.
The pair
Hamiltonian $H_{kl}$ acting on the transverse coordinates of the reggeized
gluons $\rho _{k},\rho _{k}^{\ast }$ can be written in the operator form
\cite{integrab}

\begin{equation}
H_{kl}=\ln \,\left| p_{k}\right| ^{2}+\ln \,\left| p_{l}\right| ^{2}+\frac{1%
}{p_{k}p_{l}^{\ast }}\ln \,\left| \rho _{12}\right| ^{2}\,p_{k}p_{l}^{\ast
}\,+\frac{1}{p_{k}^{\ast }p_{l}}\ln \,\left| \rho _{kl}\right|
^{2}\,p_{k}^{\ast }p_{l}-4\Psi (1)\,.
\end{equation}

Similar to the Pomeron case the BKP equation is implied to be multiplied by
a smooth function $\widetilde{\Phi }$. We will assume that this function has a
property of vanishing at small gluon momenta $\overright{p}_{r}$ ($%
r=1,2,...,n$). Similar to the Pomeron case one can verify that this
property is conserved during the BFKL evolution. Therefore we have the
freedom to add to $f$ a linear combination of functions which do not depend
on one of coordinates $\overright{\rho }_{r}$. This gives the possibility
to impose additional constraints on the solution $f$. In general, this
freedom is not enough to find a solution $f$ which vanishes at small
relative coordinates $\overright{\rho }_{rl}$. An example is the
Odderon solution constructed from three reggeized gluons:
some time ago we found a solution of the BKP equation
$\ f(\overright{\rho }%
_{1},\overright{\rho }_{2},\overright{\rho }_{3})$, which is
symmetric under the permutation $\overright{\rho }_{r}\longleftrightarrow
\overright{\rho }_{l}$ but does not vanish at $\overright{\rho }%
_{rl}\rightarrow 0$ \cite{newodd}. It is easy to see that by adding functions
which do not depend on one of coordinates $\overright{\rho }_{s}$
\beq
\varphi (\overright{\rho }_{1},\overright{\rho }_{2},%
\overright{\rho }_{3})=f(\overright{\rho }_{1},\overright{%
\rho }_{2},\overright{\rho }_{3})+\tilde{f}(\overright{\rho }_{1},%
\overright{\rho }_{2})+\tilde{f}(\overright{\rho }_{2},\overright{%
\rho }_{3})+\tilde{f}(\overright{\rho }_{3},\overright{\rho }_{2})\,,
\eeq
it is not possible to achieve
$\varphi (\overright{\rho }_{1},\overright{\rho }_{2},
\overright{\rho }_{3})=0$
at $\overright{\rho }_{r} = \overright{\rho }_{l}$,
because $f(\overright{\rho }_{r},\overright{\rho }_{r},%
\overright{\rho }_{s})$ is not symmetric to the transmutation $%
\overright{\rho }_{r}\longleftrightarrow \overright{\rho }_{s}$.

Nevertheless, let us consider the class of solutions of the BKP equation,
which are zero when one of the relative coordinates $\overright{\rho }%
_{rl}$ vanishes

\beq
\lim_{\overright{\rho }_{rl}\rightarrow 0}\,f=0\,.
\eeq
We define such solutions as belonging to the (generalized) M\"{o}bius
representation This definition is motivated by the fact, that for the
functions in the M\"{o}bius representation the pair Hamiltonians $H_{kl}$
acts in the same space of functions as in the Pomeron case, and therefore it
is M\"{o}bius invariant. \ The property
$f=0$ at $\overright{\rho} _{rl}=0$ is compatible with the BFKL evolution.

The fact, that  in the M\"{o}bius representation the space of
functions is universal, gives a
possibility to find an upper boundary for
the intercepts $\omega $ of composite states of $n$ reggeized gluons
using a variational approach (cf. \cite{GLN})

\beq
\omega \leq \frac{n(n-1)}{2}\,\omega _{BFKL}\,,
\eeq
where
\beq
\omega _{BFKL}=\bar{\alpha}_s\,4 \ln \,2\,.
\eeq
In order to obtain this bound we have used a rather rough estimate,
requiring that the average value of each pair Hamiltonian $H_{kl}$ is larger
than the minimal eigenvalue $E_{BFKL}$ of the BFKL Hamiltonian.

In the M\"{o}bius representation the Hamiltonian $H$ has the property of
holomorphic separability \cite{separab}

\beq
H=h+h^{\ast }\,,
\eeq
where

\beq
h=\sum_{1\leq k\leq l\leq n}\frac{T_{k}^{a}T_{l}^{a}}{(-N_{c})}%
h_{kl}\,,\,\,h_{kl}=\sum_{r=k,l}\left( \ln (p_{r})+\frac{1}{p_{r}}\ln (\rho
_{kl})p_{r}-\Psi (1)\right) \,.
\eeq
However, the separability does not allow to simplify the BKP\ equation,
because $h$ and $h^{\ast }$ do not commute with each other, due to the
presence of the colour matrices. Only in the multi-colour limit $%
N_{c}\rightarrow \infty $ the colour structure is drastically simplified
~\cite{integrab}. Indeed, for a general irreducible
case at $N_{c}\rightarrow \infty$ each gluon $r$ interacts with the
neighbouring gluons $r+1$ and $r-1$, and the holomorphic and
anti-holomorphic Hamiltonians $h,\,h^{\ast }\,$ commute with each other:

\beq
h=\frac{1}{2}\sum_{1\leq k\leq n}h_{k,k+1}\,,\,\,h^{\ast }=\frac{1}{2}%
\sum_{1\leq k\leq n}h_{k,k+1}\,,\,\,[h,h^*]=0\,.
\eeq
Moreover, in the multi-colour limit there are many integrals of motion $%
q_{r} $ ($r=2,3,...,n$) \cite{Baxt}

\beq
q_{r}=\sum_{1\leq i_{1}\leq i_{2}....\leq i_{r}\leq n}\rho _{i_{1}i_{2}}\rho
_{i_{2}i_{3}}...\rho
_{i_{r}i_{1}\,\,}p_{i_{1}}p_{i_{2}}...p_{i_{r}}\,,\,\,[q_{r},q_{s}]=0\,,\,%
\,[q_{r},h]=0\,,
\eeq
and the Hamiltonian $h$ coincides with the Hamiltonian for an integrable
Baxter spin model.

%%%%%%%%%%%%%%%%%%%%%%%%%%%%%%%%%%%%%%%%%%%%%%%%%%%%%%%%%%%%%%%%%%%%%%%%%
\section{Nonlinear equation for the fan diagrams}

Let us now write down the evolution equation which sums the fan diagrams.
To be definite, we consider a simplified model ~\cite{BRV} of the elastic
scattering of two quark-antiquark pairs (Fig.1):
\begin{figure}
\begin{center}
\epsfig{file=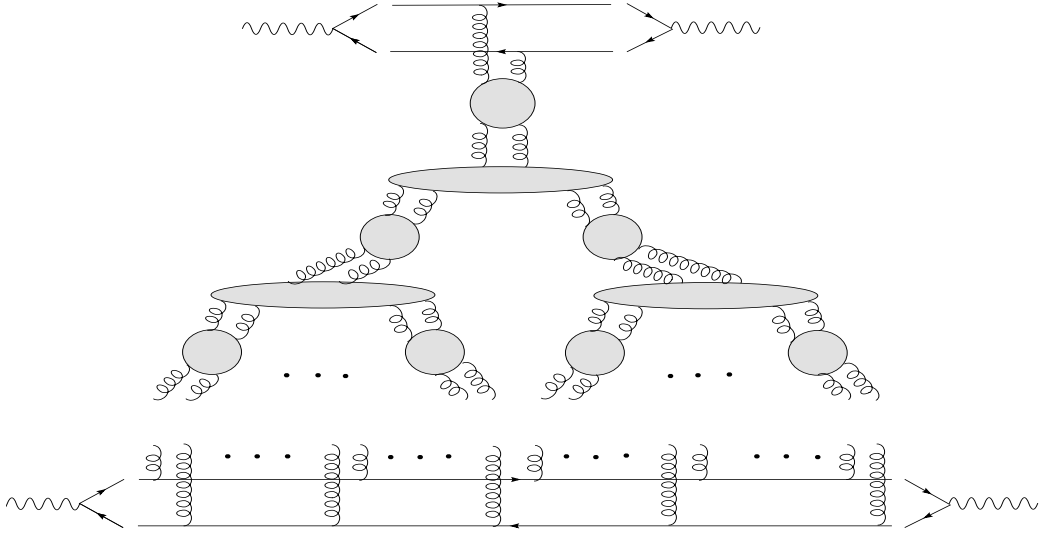,width=140mm,height=7cm}
\caption{The fan diagram equation 
(the coupling of gluons to quark lines 
includes a sum over all possibilities).}
\end{center}
\label{fanfig}
\end{figure}
the upper (smaller) quark
pair couples to a single BFKL ladder
\footnote{Note that, because of the structure of the BFKL kernel and of the
reggeization of the gluon, the coupling of
a single BFKL Green's function to the $q\bar{q}$ system contains an arbitrary
number of elementary gluon propagators being attached to the quark lines.},
whereas the lower (larger) quark pair
couples to an arbitrary number of BFKL Pomerons (Fig.1).
Both couplings are taken to be of the eikonal type. As a consequence,
at the lower quark pair the couplings of the BFKL Green's functions factorize.
When summing the fan diagrams, the transverse coordinates
of the lower quark pair, $\frho_0$, $\frho_0'$, are kept fixed.
If $\Psi(\frho_1,\frho_2;\frho_0, \frho_0';Y)$ denotes the non-amputated
two-gluon amplitude, the equation has the form:
\begin{eqnarray}
\frac{\partial \Psi(\frho_1,\frho_2;\frho_0, \frho_0';Y)}{\partial Y}=
-\frac{\bar{\alpha}_s}{2}
(H_{12} \Psi)(\frho_1,\frho_2;\frho_0, \frho_0',Y)\hspace{2cm}\nonumber \\
- \bar{\alpha}_s^2  \int d^2 \rho_{\alpha} d^2 \rho_{\alpha'}
       d^2 \rho_{\beta} d^2 \rho_{\beta'}
{\cal V} (\frho_1,\frho_2;\frho_{\alpha},\frho_{\alpha'},
         \frho_{\beta}, \frho_{\beta'})
         \Psi(\frho_{\alpha},\frho_{\alpha'};\frho_0, \frho_0';Y)
\Psi(\frho_{\beta},\frho_{\beta'};\frho_0, \frho_0';Y)\,,
\label{fansum}
\end{eqnarray}
where $H_{12}$ is the BFKL Hamiltonian, and
${\cal V} (\frho_1,\frho_2;\frho_{\alpha},\frho_{\alpha'},
         \frho_{\beta}, \frho_{\beta'})$
denotes the conformal invariant $2 \to 4$ transition vertex of reggeized
gluons \cite{Bartels:1994jj,Bartels:1995kf,Braun:1997nu}.
When supplemented with the initial condition:
\beq
\Psi(\frho_1,\frho_2;\frho_0, \frho_0';Y=0)
= \alpha_s\; \left(
f^{(0)}(\frho_1,\frho_2;\frho_0,\frho_0')
+ f^{(0)}(\frho_1,\frho_2;\frho_0',\frho_0) \right) \, ,
\eeq
where $f^{(0)}$ is proportional to the two gluon propagator in the
M\"obius representation,
this nonlinear equation sums the fan diagrams coupled to the lower
quark-antiquark pair. In order to obtain a physical scattering amplitude, we
multiply $\Psi$ with suitable wave functions of external particles
and integrate over the transverse coordinates
$\frho_1$, $\frho_2$, $\frho_0$, $\frho_0'$.

In the momentum space the $2 \to 4$ transition vertex was found
~\cite{Bartels:1994jj} to consist of three pieces:
\beq  \bar{\alpha}_s^2 {\cal V} (\bq_1,\bq_2;\bk_1,\bk_2,\bk_3,\bk_4) =
\delta_{a_1 a_2} \delta_{a_3 a_4} V(1234)
     + \delta_{a_1 a_3} \delta_{a_2 a_4} V(1324)
     + \delta_{a_1 a_4} \delta_{a_2 a_3} V(1423)\,,
\label{fullvertex}
\eeq
where we have introduced the short-hand notation
\beq
V(1234) = V(\bq_1,\bq_2;\bk_1,\bk_2,\bk_3,\bk_4),
\eeq
and the subscripts $a_i$ refer to the color degrees of the reggeized gluons.
Obviously, the vertex $\cal V$ is completely symmetric under the
exchange of any two gluon lines $i$ and $j$ $(i,j=1,2,3,4)$. Furthermore,
the function $V(1234)$ vanishes if one of the four momenta $\bk_i$ goes to
zero. A convenient representation is the following:
\[
V(1234)D_2 =\frac{1}{2}g^2 \Bigl[ G(1,2+3,4)+G(2,1+3,4)+G(1,2+4,3)+G(2,1+4,3)
\]
\begin{equation}
- G(1+2,3,4)-G(1+2,4,3)-G(1,2,3+4)-G(2,1,3+4)+G(1+2,0,3+4) \Bigr] \, .
\label{verbart}
\end{equation}
The function $G(1,2,3)$ is the non-forward
extension~\cite{Braun:1997nu,Vacca:1998kc} of the $G$-function
introduced in \cite{Bartels:1994jj}. It acts on the (amputated) 2-gluon test
functions in the M\"obius representation
$D_2(\bq_1,\bq_2)$, and it consists of two pieces
\begin{equation}
G(\mbox{\boldmath $k$}_1,\mbox{\boldmath $k$}_2,\mbox{\boldmath $k$}_3)=G_1(%
\mbox{\boldmath $k$}_1,\mbox{\boldmath $k$}_2,\mbox{\boldmath $k$}_3)+G_2(%
\mbox{\boldmath $k$}_1,\mbox{\boldmath $k$}_2,\mbox{\boldmath $k$}_3) \, ,
\label{Gfunction}
\end{equation}
The first term, $G_1$, belongs to the diagrams describing the
emission of a real gluon,
\[
G_1(\mbox{\boldmath $k$}_1,\mbox{\boldmath $k$}_2,\mbox{\boldmath $k$}_3)=
g^2N_c\,\int \frac{d^2\mbox{\boldmath $q$}_1d^2\mbox{\boldmath $q$}_2}{%
(2\pi)^3} \delta^2 (\mbox{\boldmath $q$}_1+\mbox{\boldmath $q$}_2-%
\mbox{\boldmath $k$}_1-\mbox{\boldmath $k$}_2-\mbox{\boldmath $k$}_3)D_2(%
\mbox{\boldmath $q$}_1,\mbox{\boldmath $q$}_2)
\]
\begin{equation}
\left(\frac{(\mbox{\boldmath $k$}_2+\mbox{\boldmath $k$}_3)^2}{(%
\mbox{\boldmath $q$}_1-\mbox{\boldmath $k$}_1)^2 \mbox{\boldmath $q$}_2^2} +%
\frac{(\mbox{\boldmath $k$}_1+\mbox{\boldmath $k$}_2)^2}{\mbox{\boldmath $q$}%
_1^2(\mbox{\boldmath $q$}_2-\mbox{\boldmath $k$}_3)^2}-\frac{%
\mbox{\boldmath
$k$}_2^2}{(\mbox{\boldmath $q$}_1-\mbox{\boldmath $k$}_1)^2 (%
\mbox{\boldmath
$q$}_2-\mbox{\boldmath $k$}_3)^2}-\frac{(\mbox{\boldmath $k$}_1+%
\mbox{\boldmath $k$}_2+\mbox{\boldmath $k$}_3)^2}{\mbox{\boldmath $q$}_1^2 %
\mbox{\boldmath $q$}_2^2}\right),  \label{proper}
\end{equation}
whereas the second part is related to the virtual correction present in the
reggeized gluon trajectory:
\begin{eqnarray}
G_2(\mbox{\boldmath $k$}_1,\mbox{\boldmath $k$}_2,\mbox{\boldmath $k$}%
_3)&=&
- \left[\omega(\mbox{\boldmath $k$}_2)-\omega(\mbox{\boldmath $k$}_2+%
\mbox{\boldmath $k$}_3)\right]
D_2(\mbox{\boldmath $k$}_1,\mbox{\boldmath $k$}_2+\mbox{\boldmath $k$}%
_3) \nonumber \\
&& -\left[\omega(\mbox{\boldmath $k$}_2)-\omega(\mbox{\boldmath $k$}_1+%
\mbox{\boldmath $k$}_2)\right] D_2(\mbox{\boldmath $k$}_1+
\mbox{\boldmath $k$}_2,\mbox{\boldmath $k$}_3)\,.  %\label{improper}
\end{eqnarray}

The function $G(\mbox{\boldmath $k$}_1,\mbox{\boldmath $k$}_2,
\mbox{\boldmath $k$}_3)$ has the property to be zero for
$\mbox{\boldmath $k$}_1=\mbox{\boldmath $0$}$ or
$\mbox{\boldmath $k$}_3=\mbox{\boldmath $0$}$
(but not for $\mbox{\boldmath $k$}_2=\mbox{\boldmath $0$}$), so that one may
easily see that the vertex $V(\mbox{\boldmath $k$}_1,\mbox{\boldmath $k$}_2,%
\mbox{\boldmath $k$}_3,\mbox{\boldmath $k$}_4) \rightarrow 0$ for any $%
\mbox{\boldmath $k$}_i \rightarrow \mbox{\boldmath $0$}$.
This relation must be satisfied by any gauge invariant description of a
t-channel 4-gluon state coupled to a colorless scattering projectile.

The expression in the coordinate representation was given in
\cite{Braun:1997nu,Vacca:1998kc} and can
be written in terms of two non-local operators, $A_1$ and $A_2$.
The operator $A_1$ is
defined as follows:

\begin{equation}
G_1(\mbox{\boldmath $r$}_1,\mbox{\boldmath $r$}_2,\mbox{\boldmath $r$}%
_3)=A_1 D_2(\mbox{\boldmath $r$}_1,\mbox{\boldmath $r$}_3),
\end{equation}
and it has the following form:
\[
A_1=\frac{g^2N_c}{8\pi^3}\Bigl[ 2\pi\delta^2(\mbox{\boldmath $r$}%
_{23})\partial_3^2(c-\ln r_{13})\partial_3^{-2}+ 2\pi\delta^2(%
\mbox{\boldmath $r$}_{12})\partial_1^2(c-\ln r_{13})\partial_1^{-2}
\]
\begin{equation}
-2\frac{{\bf r}_{12}{\bf r}_{23}}{r_{12}^2 r_{23}^2}- 2\pi (c-\ln
r_{13})(\delta^2(\mbox{\boldmath $r$}_{12})+\delta^2(\mbox{\boldmath $r$}%
_{23})) - 4\pi^2\delta^2(\mbox{\boldmath $r$}_{12})\delta^2(%
\mbox{\boldmath
$r$}_{23}) (\mbox{\boldmath $\partial$}_1+\mbox{\boldmath $\partial$}_3)^2
\partial_1^{-2} \partial_3^{-2}\Bigr]\,.  \label{a1}
\end{equation}
Here $r_{ij}=|\mbox{\boldmath $r$}_{ij}|$, $\partial_i=|%
\mbox{\boldmath
$\partial$}_i|$ and $c=\ln (2/m)+\psi(1)$, and $m$ is a gluon mass which
provides an infrared regulator.

In order to transform $G_2$ to coordinate space, an ultraviolet
regularization (with a parameter $%
\epsilon$) is necessary due to the presence of the gluon trajectory terms.
The dependence on this regularization will disappear at the end. One obtains
\begin{equation}
G_2(\mbox{\boldmath $r$}_1,\mbox{\boldmath $r$}_2,\mbox{\boldmath $r$}%
_3)=A_2 D_2(\mbox{\boldmath $r$}_1,\mbox{\boldmath $r$}_3)\,,
\end{equation}
where the operator $A_2$ is given by
\begin{eqnarray}
A_2&=&-\frac{g^2N_c}{8\pi^3}\Bigl[\frac{1}{r_{23}^2}-2\pi c\delta^2(%
\mbox{\boldmath $r$}_{23}) \Bigr] +\delta^2(\mbox{\boldmath $r$}%
_{23})\omega(-i\partial_3)  \nonumber \\
&& -\frac{g^2N_c}{8\pi^3}\Bigl[\frac{1}{r_{12}^2}-2\pi c\delta^2(%
\mbox{\boldmath $r$}_{12}) \Bigr] +\delta^2(\mbox{\boldmath $r$}%
_{12})\omega(-i\partial_1)\,.  \label{a2}
\end{eqnarray}
For the singular operators $1/r_{12}^2$ and $1/r_{23}^2$ one may use the
ultraviolet regularization
\begin{equation}
\frac{1}{r^2}\equiv\frac{1}{r^2+\epsilon^2}+2\pi\delta^2(r)\ln\epsilon \, 
\label{regul}
\end{equation}
with the understanding that $\epsilon \to 0$ at the end of the calculation.
In the sum of the two operators, $A=A_{1}+A_{2}$, the terms containing
$\ln m$ cancel, thus the
dependence on the gluon mass disappears, and $G(\mbox{\boldmath $r$}_{1},%
\mbox{\boldmath $r$}_{2},\mbox{\boldmath $r$}_{3})=AD_{2}(%
\mbox{\boldmath
$r$}_{1},\mbox{\boldmath $r$}_{3})$ is infrared stable.

\section{M\"{o}bius representation for the fan equation}

Let us now compare the fan diagram equation (\ref{fansum})
with the Balitsky-Kovchegov
equation (BK-equation) ~\cite{BalKov}:
\begin{equation}
\frac{d}{dY}N_{\overright{x},\overright{y}}=\bar{\alpha}_s
\int \frac{d^{2}z}{2\pi }\,\frac{\left| x-y\right| ^{2}}{\left|
x-z\right| ^{2}\left| y-z\right| ^{2}}\,\left( N_{\overright{x},%
\overright{z}}+N_{\overright{y},\overright{z}}-N_{%
\overright{x},\overright{y}}-N_{\overright{x},\overright{%
z}}N_{\overright{y},\overright{z}}\right) \,.
\label{BK}
\end{equation}
We will show
that, by taking $N_c$ to be large and restricting ourselves to functions
in the M\"obius representation, the nonlinear fan
diagram equation (\ref{fansum}) coincides with the BK equation.

Beginning with the linear part of (\ref{fansum}) which has been discussed in section 1,
we make use of the freedom to add to the $\Psi(\frho_1,\frho_2)$ \footnote
{From now on, for the function $\Psi(\frho_1,\frho_2;\frho_0,\frho_0';Y)$ 
we will simply write $\Psi(\frho_1,\frho_2)$.}  
(which in the dipole approach is a symmetric function)
a new function which depends only on one of the two coordinates. Moreover
we scale the result by a factor proportional to $\alpha_s$. We choose
\beq
\tilde{\Psi}(\frho_1,\frho_2) =
B \left( \Psi(\frho_1,\frho_2) - \frac{1}{2}\Psi(\frho_1,\frho_1)
- \frac{1}{2}\Psi(\frho_2,\frho_2)\right) \,.
\label{shift}
\eeq
With this choice we have
\beq
\tilde{\Psi}(\frho,\frho) = 0,
\label{cond_ct}
\eeq
i.e. $\tilde{\Psi}$ is in the M\"obius representation. Later on, we will
identify $\tilde{\Psi}$ with the dipole distribution, $N$, and we will 
determine the constant $B$.
The condition (\ref{cond_ct}) is
the color transparency relation (CTR). We remind that the shift (\ref{shift})
is allowed because of the `good' properties of the impact factor
which vanishes when either $\bq_1=0$ or $\bq_2=0$.

Let us now turn to the non-linear term in (\ref{fansum}).
As mentioned before, the
$2 \to 4$ gluon vertex is zero, when one of momenta $k_{i}$ tends to zero at
fixed $q_{1}$ and $q_{2}$. This means that after performing the Fourier
transformation of the equation and switching from the momenta $\bk_i$ ($i=1,...,4$) to
the coordinates $\frho_{\alpha}$, $\frho_{\alpha'}$,
$\frho_{\beta}$, and $\frho_{\beta'}$ we are, again, allowed to add
contributions to $\Psi(\frho_{\alpha}, \frho_{\alpha'})$ which
lead to the condition $\tilde{\Psi}(\frho,\frho) = 0$.
They can be described by the projector $\theta^{UV}$ used in eq.
(\ref{subtraction}).
As a result, we have rewritten the fan diagram equation for $\Psi$ into
an equation for $\tilde{\Psi}$ which belongs to the M\"obius space of
functions.

The final step now is the observation that, when projecting on color singlet
states in the (12) and (34) subsystems,
for large $N_c$, only the first term of (\ref{fullvertex}), $V(1234)$,
contributes.
When acting on functions
$\tilde{\Psi}(\frho_1,\frho_2)$ and $\tilde{\Psi}(\frho_3,\frho_4)$
which are in the M\"obius representation,
the second line of (\ref{verbart}) does not
contribute.
For the remaining terms of $V(1234)$, the sum of the two operators
$A_1$ and $A_2$ becomes simply
\beq
- 4 \frac{g^2}{2} \frac{g^2 N_c}{8\pi^3}
\frac{\rho_{12}^2}{\rho_{13}^2 \rho_{23}^2}.
\eeq
which coincides with the nonlinear term in the BK-equation,
if we choose in eq. (\ref{shift})
\beq
B=8\pi \bar{\alpha}_s.
\eeq
Therefore, when identifying $N$ with our subtracted function (\ref{shift}),
$\tilde{\Psi}$, the Balitsky-Kovchegov equation follows from the fan diagram
equation, provided we restrict ourselves to the leading term at large $N_c$.

Next one may ask what kind of contribution is given by those terms
in ${\cal V} (\bq_1,\bq_2;\bk_1,\bk_2,\bk_3,\bk_4)$, eq. (\ref{fullvertex}),
that we have neglected so far. In order to do that let us consider, inside
a fan diagram, the splitting from a $\tilde\Psi'$
state to a $\tilde{\Psi} \tilde{\Psi}$ state (we imagine that the
subtraction $\Psi \to \tilde{\Psi}$ which guarantees
$\tilde{\Psi}(\frho,\frho)=0$ has already been done). From the
calculations shown in the appendix we
derive the contribution:
\begin{eqnarray}
\frac{2}{N_c^2-1} \!\!\!\!\!\!\!&&\!\!\!\!\!\!
\int d^2 \rho_1 d^2\rho_2 \frac{\alpha_s^2 N_c}{\pi} \Biggl\{
- 2 \frac{\tilde{\Psi}'(\frho_1,\frho_2)}{|\frho_{12}|^4} \int d^2\rho_3
\frac{|\rho_{12}|^2}{|\rho_{13}|^2|\rho_{32}|^2}
\tilde{\Psi}(\frho_1,\frho_3) \tilde{\Psi}(\frho_3,\frho_2)- \nonumber \\
&+& \frac{\tilde{\Psi}'(\frho_1,\frho_2)}{|\frho_{12}|^4}
\left[-\pi H_{12}\tilde{\Psi}(\frho_1,\frho_2)\right]
\tilde{\Psi}(\frho_1,\frho_2)+
\tilde{\Psi}(\frho_1,\frho_2)\left[-\pi H_{12}
\tilde{\Psi}(\frho_1,\frho_2)\right]+\nonumber\\
&+&\left[\frac{1}{|\frho_{12}|^4}\pi H_{12}
\tilde{\Psi}'(\frho_1,\frho_2)\right]
\tilde{\Psi}(\frho_1,\frho_2)\tilde{\Psi}(\frho_1,\frho_2)
\Biggr\} \, .
\end{eqnarray}
It is now crucial to recall the hermitian
symmetry of the BFKL Hamiltonian
$H_{12}$ for the last term, according to
\beq
\int d^2 \rho_1 d^2\rho_2
\left[\frac{1}{|\frho_{12}|^4}\pi H_{12}\tilde{\Psi}'(\frho_1,\frho_2)\right]
\tilde{\Psi}^2(\frho_1,\frho_2)=
\int d^2 \rho_1 d^2\rho_2
\frac{\tilde{\Psi}'(\frho_1,\frho_2)}{|\frho_{12}|^4}
\pi H_{12} \tilde{\Psi}^2(\frho_1,\frho_2) \,.
\eeq
Performing the scaling by the factor $B$ which takes us from
$\tilde{\Psi}$ to the dipole distribution $N_{\overright{x},\overright{y}}$
we find the simple form
\begin{eqnarray}
\frac{1}{2}\frac{1}{N_c^2-1}\!\!\!\! &\bar{\alpha}_s&\!\!\!\!
\int \frac{d^{2}z}{2\pi }\,\frac{\left| x-y\right| ^{2}}{\left|
x-z\right| ^{2}\left| y-z\right| ^{2}}\,\Biggl[
-2 N_{\overright{x},\overright{z}}N_{\overright{z},\overright{y}}+\nonumber\\
&+& 2\left(
N_{\overright{x},\overright{z}}+N_{\overright{z},\overright{y}}-
N_{\overright{x},\overright{y}}\right) N_{\overright{x},\overright{y}}
- \left(N^2_{\overright{x},\overright{z}}+N^2_{\overright{z},\overright{y}}-
N^2_{\overright{x},\overright{y}}\right)=\nonumber \\
&=& \bar{\alpha}_s
\int \frac{d^{2}z}{2\pi }\,\frac{\left| x-y\right| ^{2}}{\left|
x-z\right| ^{2}\left| y-z\right| ^{2}}\,\Biggl[
-\frac{1}{2}\frac{1}{N_c^2-1}
\left( N_{\overright{x},\overright{z}}+N_{\overright{z},\overright{y}}-
N_{\overright{x},\overright{y}}\right)^2\Biggr]
\end{eqnarray}

The negative sign indicates that these large-$N_C$ corrections to the
triple Pomeron vertex again lead to the saturation for
evolution in rapidity.
The factor $1/[2(N_c^2-1)]=1/16$ seems to suggest that this
contribution should not play a crucial role. Nevertheless a direct
investigation would be interesting.

It is important to note that, when going beyond the large
$N_c$ limit, there are other corrections which slightly complicate
the simple structure
of the nonlinear fan diagram equation. They are due to the evolution
of the colourless state of $2n$ reggeized gluons with $n>1$:
for example, in leading order $1/N_c$, the four-gluon state consists of two
noninterating Pomeron states. Each interaction
between the Pomerons costs a suppression of the order $1/(N_c^2-1)$, i.e.
it is of the same order as the corrections to the triple Pomeron vertex
discussed above.
Therefore, a consistent treatment of corrections beyond the large-$N_c$ limit
has to include these corrections to the Hamiltonian of the evolution of four
gluon state.
As a first step, one can replace the single nonlinear evolution equation
for $\tilde{\Psi}$ by a system of coupled
equations, which describe the evolution of
the two-gluon amplitude $\tilde{\Psi}$ and of the four-gluon Green's function 
$G_4$. The equation for $N$ reads:
\begin{eqnarray}
\frac{d}{dY}N_{\overright{x},\overright{y}}=\bar{\alpha}_s
\int \frac{d^{2}z}{2\pi }\,\frac{\left| x-y\right| ^{2}}{\left|
x-z\right| ^{2}\left| y-z\right| ^{2}}\,\left( N_{\overright{x},%
\overright{z}}+N_{\overright{y},\overright{z}}-N_{%
\overright{x},\overright{y}}-
N_4(\bx,\bz;\by,\bz;Y)
\right.\nonumber \\ \left.
-\frac{1}{2}\frac{1}{N_c^2-1}
\left( N_{\overright{x},\overright{z}}+N_{\overright{z},\overright{y}}-
N_{\overright{x},\overright{y}}\right)^2 \right)
\label{BKext}
\end{eqnarray}
The argument structure of $N_4(\frho_1,\frho_2;\frho_3,\frho_4;Y)$
indicates that the first pair of gluons at positions $\frho_1$, $\frho_2$
are in a color singlet; the same applies to the second pair at
$\frho_3$, $\frho_4$.
In leading order $1/N_c$, $N_4(\frho_1,\frho_2;\frho_3,\frho_4;Y)$
equals the product $N_{\frho_1, \frho_2} N_{\frho_3, \frho_4}$.
To include the first correction we write 
\beq
N_4(\frho_1,\frho_2;\frho_3,\frho_4;Y)\;=\;
N_{\frho_1,\frho_2} N_{\frho_3, \frho_4} +
\Delta N_4 (\frho_1,\frho_2;\frho_3,\frho_4;Y)
\eeq   
A second equation for $\Delta N_4$ describes the evolution of the four gluon 
Green's function where the first interaction of the order $1/(N_c^2-1)$ between
the two dipole cross sections $N$ is kept:
\begin{eqnarray}
\frac{d}{dY} \Delta N_4 (\frho_1,\frho_2;\frho_3,\frho_4;Y) 
&=& - \frac{\bar{\alpha}_s}{2(N_c^2 -1)} \left(H_{12} + H_{34} \right) 
\left(N_{\frho_1,\frho_3} N_{\frho_2, \frho_4} + 
N_{\frho_1,\frho_4} N_{\frho_2, \frho_3} \right) \nonumber \\
&- & 
\frac{\bar{\alpha}_s}{2} \left(H_{12} + H_{34} \right)
\Delta N_4 (\frho_1,\frho_2;\frho_3,\frho_4;Y)  
\end{eqnarray}
When combined with the integral kernel in eq.(\ref{BKext}), 
$\Delta N_4$ can be interpreted as a $O(1/N_C^2)$ loop-correction to the  
triple Pomeron vertex in the space of M\"obius functions. It would be 
interesting to study further correction terms of higher order in $O(1/N_C^2)$. 
%%%%%%%%%%%%%%%%%%%%%%%%%%%%%%%%%%%%%%%%%%%%%%%%%%%%%%%%%%%%%%%%%%%%%%%%
\section{Conclusions}
In this paper we have investigated some consequences of the M\"obius
invariance of the BFKL Hamiltonian. When combined with the fact that
Green's functions of reggeized gluons couple to impact factors of colorless
external states, this invariance allows to redefine the two-gluon Green's
function in such a way that it vanishes as the two coordinates of the
gluons coincide. This property defines what we have named the `M\"obius
representation'. For the triple Pomeron vertex we have shown that
this M\"obius representation leads to a very simple form of the
interaction kernel.

The use of the M\"obius representation also allows to study the connection
between the reggeon calculus in QCD (formulated in terms of t-channel
partial waves) and the dipole picture. The latter is now widely been
used for studies of, for example, saturation phenomena in deep inelastic 
scattering and in heavy ion collisions. 
An advantage of starting from the reggeon approach
lies in the fact that it allows to go beyond the LO approximation 
and beyond the large-$N_c$ limit.
As an example, we have studied the fan diagram equation. In the large-$N_c$
limit it coincides with the BK equation. We then have computed
the $1/N_c^2$ suppressed corrections to the triple Pomeron vertex
which are not contained in the BK equation. Accuracy of the order 
$1/N_c^2$ requires to consider also corrections in the evolution of the four 
gluon states; we propose a first modification of the BK equation which 
includes these corrections.   

A few years ago the next-to-leading
corrections (NLO) to the BFKL equation have been calculated in the framework 
of the reggeon approach \cite{Next}. Therefore it is natural to
use these results in the M\"{o}bius representation and to 
study their role in the dipole picture.
%%%%%%%%%%%%%%%%%%%%%%%%%%%%%%%%%%%%%%%%%%%%%%%%%%%%%%%%%%%%%%%%%%%%%%%%
\section{Acknowledgments}
G.P. Vacca and L.N. Lipatov wish to thank the II.Instit\"ut f\"ur
Theoretische Physik, University Hamburg, and DESY for the warm hospitality.
%%%%%%%%%%%%%%%%%%%%%%%%%%%%%%%%%%%%%%%%%%%%%%%%%%%%%%%%%%%%%%%%%%%%%%%%
\section*{Appendix: The non planar contribution of the triple Pomeron vertex}
Let us recall the structure of the two non-planar contributions
to the vertex (\ref{fullvertex}), $V(1324)$ and $V(1423)$.
They have been studied previously in the context of the coupling of
three pomeron states with definite conformal
weights~\cite{Lotter:1996vk,Korchemsky:1997fy,Bialas:1997xp}.

Each of them gives the same contribution, which can be derived from $V(1324)$:
\[
V(1324)D_2 =\frac{1}{2}g^2 \Bigl[ G(1,3+2,4)+G(3,1+2,4)+G(1,3+4,2)+G(3,1+4,2)
\]
\begin{equation}
- G(1+3,2,4)-G(1+3,4,2)-G(1,3,2+4)-G(3,1,2+4)+G(1+3,0,2+4) \Bigr] \,.
\label{verbart2}
\end{equation}
First we note that, when coupling to color singlet states in the 
systems (12) and (34), these contributions are color suppressed by a
factor $1/(N_c^2-1)$.
Moreover one can immediately see that, when considering, due to the
gauge freedom, functions $\tilde{\Psi}(\frho_1,\frho_2)$ in the
M\"obius representation, the second and third terms give no
contribution.
In the first and the fourth terms almost all the pieces cancel in the
M\"obius space of functions, and one is left with only two terms which 
in coordinate space have the form:
\beq
-\frac{\alpha_s^2 N_c}{\pi(N_c^2-1)} \left( \delta^{(2)}(\frho_{23})
\frac{|\rho_{14}|^2}{|\rho_{13}|^2|\rho_{34}|^2}+
\delta^{(2)}(\frho_{14})
\frac{|\rho_{13}|^2}{|\rho_{12}|^2|\rho_{23}|^2}\right)
\eeq
When integrated with the $\tilde{\Psi}$ states. both pieces give the same 
contribution.

As to the last term in (\ref{verbart2}), when looking at the structure of 
$G$ it can be seen that the last
term simply corresponds to a BFKL kernel acting on
the amputated function $D_2$. However, in our fan resummation 
the vertex acts on non-amputated functions, and the term can be written as:
\beq
\frac{\alpha_s^2 N_c}{\pi(N_c^2-1)}  \delta^{(2)}(\frho_{13})
\delta^{(2)}(\frho_{24}) \frac{1}{|\rho_{12}|^4} \pi H_{12}
\eeq
which acts on the single two gluon state before the splitting.

Let us finally consider the remaining four terms  of (\ref{verbart2}) 
(the first four terms in the second line).
Due to the fact that $G$ is an operator acting
on the two gluon state (and not on the four gluon state), it is convenient to 
use an expression different from the previous form of the operators 
$A_1$ and $A_2$.

Namely, a direct investigation of, for example, $-G(1+3,2,4)$ gives:
\beq
\frac{\alpha_s^2 N_c}{\pi(N_c^2-1)} \delta^{(2)}(\frho_{13})
\left[ \frac{|\rho_{14}|^2}{|\rho_{12}|^2|\rho_{24}|^2}
-\pi \log{|\rho_{14}|^2} \delta^{(2)}(\frho_{34}) \right]=
\frac{\alpha_s^2 N_c}{\pi(N_c^2-1)} \delta^{(2)}(\frho_{13}) Z^1_{4,2}.
\eeq
Therefore it is easy to see that the action of this part of the
non planar vertex reads
\begin{eqnarray}
\int d^2 \rho_1 d^2\rho_4 \frac{\alpha_s^2 N_c}{\pi(N_c^2-1)}
\frac{\bar{\Psi}'(\frho_1,\frho_4)}{|\rho_{14}|^4}
\!\!\!\!\!\!\!\!\!\!\!\!\!\! && \Biggl\{
\int d^2\rho_3 \left[ Z^1_{4,3} \tilde{\Psi}(\frho_1,\frho_3)
+ Z^4_{1,3} \tilde{\Psi}(\frho_3,\frho_4)\right] \tilde{\Psi}(\frho_1,\frho_4)
+\nonumber \\
&+&\!\!\!\! \tilde{\Psi}(\frho_1,\frho_4) \int d^2\rho_3 \left[ Z^1_{4,3}
\tilde{\Psi}(\frho_1,\frho_3)
+ Z^4_{1,3} \tilde{\Psi}(\frho_3,\frho_4)\right]
\Biggr\}\,.
\end{eqnarray}
The integral operator depending on the kernel $Z^i_{j,k}$ in the two
lines are simply BFKL kernel operators and can be written as 
\beq
\int d^2\rho_3 \left[ Z^1_{4,3} \tilde{\Psi}(\frho_1,\frho_3)
+ Z^4_{1,3} \tilde{\Psi}(\frho_3,\frho_4)\right] =
-\pi H_{14}  \tilde{\Psi}(\frho_1,\frho_4)\,.
\eeq
This form follows from the relation (\ref{dipole_real}).
We therefore conclude that the non planar vertex can be written in terms of 
well known objects.

%%%%%%%%%%%%%%%%%%%%%%%%%%%%%%%%%%%%%%%%%%%%%%%%%%%%%%%%%%%%%%%%%%%%%%%%%%%

\end{document}